# Improving Quality of Service and Reducing Power Consumption with WAN accelerator in Cloud Computing Environments

Shin-ichi Kuribayashi[1]

[1]Department of Computer and Information Science, Seikei University, Japan

E-mail: `kuribayashi@.st.seikei.ac.jp`

*Abstract*

*The widespread use of cloud computing services is expected to deteriorate a Quality of Service and toincrease the power consumption of ICT devices, since the distance to a server becomes longer than before. Migration of virtual machines over a wide area can solve many problems such as load balancing and power saving in cloud computing environments.*

*This paper proposes to dynamically apply WAN accelerator within the network when a virtual machine is moved to a distant center, in order to prevent the degradation in performance after live migration of virtual machines over a wide area. mSCTP-based data transfer using different TCP connections before and after migration is proposed in order to use a currently available WAN accelerator. This paper does not consider the performance degradation of live migration itself. Then, this paper proposes to reduce the power consumption of ICT devices, which consists of installing WAN accelerators as part of cloud resources actively and increasing the packet transfer rate of communication link temporarily. It is demonstrated that the power consumption with WAN accelerator could be reduced to one-tenth of that without WAN accelerator.*

## 1. Introduction

Cloud computing services are rapidly gaining in popularity [1]-[3], which enable users to access huge computing resources (processing ability, storage, etc.) distributed over the network from any terminals for a required length of time without any need to worry about where resources are located or how resources are structured internally. The background to this rapid penetration includes the availability of high-speed networks, and the development of virtual computing, grid computing and other advanced computing technologies. In recent years, enterprises are shifting from building their own information systems to using cloud computing services increasingly, because cloud computing services are easy to use and enable them to reduce both their business cost and environmental impacts.

As a lot of servers and storages will be centralized in cloud computing environments, it is necessary to deal with failures and to balance the processing load. It is also required to decrease the power consumption of ICT devices.

In cloud computing environments, migration of virtual machines (VMs) over a wide area can solve many problems such as load balancing and power saving. Migration technology is used to move the memory spaces of VMs from one physical server to a different physical server while ensuring service continuity. In particular, live migration is intended to move VMs with virtually no disruption to the services being provided. There are studies that assume that VMs are moved over a wide area rather than keeping them confined in the same site[4]-[13]. Such wide-area migration will improve robustness against wide-area disasters and the effectiveness of load balancing. When a VM is moved to a distant center, the performance may deteriorate (e.g, slow response and decreased throughput) due to an increase in network delay or a reduction in bandwidths. It is required to prevent degradation in performance after live migration of VMs over a wide area. We propose to automatically apply WAN accelerator[14]-[18] (also known as WAN optimization)to prevent degradation in performance when the network delay between the terminal and the center exceeds a certain threshold as a result of moving a VM.

Although ICT makes it possible to reduce power consumption in the country by optimizing physical distribution, optimizing production, and reducing human movements (commuting and business trips) [19],[20], it is required to make efforts to prevent an increase in power consumption of ICT devices as much as possible. Cloud computing, which uses huge computing and network resources, are naturally subject to such efforts. To reduce the power consumed by the entire ICT devices, it is necessary to take coordinated measures to implement power saving in data centers, communication networks and power networks, instead of seeking to save power in each of these independently. The authors have proposed a guideline and a procedure to exchange information needed for this coordination [21]-[23]. With a view to further reducing the power consumed by the entire ICT devices, we present a new power saving measure with WAN accelerators in this paper. Although WAN accelerators have been introduced to prevent degradation in performance caused by a long network delay in WANs originally, they can also dramatically shorten communication time of applications such a file transfer that transfer a huge volume of data continuously but do not require real-time transfer. The reduction in communication time can also reduce the power used by data centers and network devices. Therefore, the introduction of WAN accelerators as part of cloud resources is useful not only for reducing communication time but also for reducing the power consumption by ICT devices.

The rest of this paper is organized as follows. Section 2 proposes to dynamically apply WAN accelerator within the network to prevent the degradation in performance after live migration of VMs over a wide area. This paper does not consider the performance degradation of live migration itself. Section 3 presents a method of reducing the power consumption of the entire ICT devices in a cloud computing environment, which consists of installing WAN accelerators as part of cloud resources actively and increasing the packet transfer rate of communication link temporarily. Section 4 explains the related work. Finally, Section 5gives the conclusions. This paper is an extension of the study in Reference [18].

## 2. Method of preventing the degradation in performance after VM live migration over a wide area

### 2.1 Overview[18]

A lot of servers and storages will be centralized in cloud computing environments. While this approach may bring economic benefits, it may sacrifice communication performance and usability for users because it will increase network delay time and traffic congestion. WAN accelerators, which aim to accelerate a broad range of applications and protocols over a WAN, are widely introduced to cope with this type of problem. For example, in the case where TCP is used, the introduction of a WAN accelerator allows the use of ACK proxy responses, the expansion of the TCP window size, and slow-start control[14]. These features can improve response time and enhance throughput. The use of data compression and caching is also useful because they can reduce both the volume of traffic and the bandwidths used in the network.

International Journal of Computer Networks & Communications (IJCNC) Vol.5, No.1, January 2013

When a VM is moved to a distant center, the performance may deteriorate (e.g, slow response and decreased throughput) due to an increase in network delay or a reduction in bandwidths. As a measure to prevent this degradation, this paper proposes to apply WAN accelerator in the network automatically whenever network delay has increased or the bandwidths available have decreased beyond certain specified levels. The proposed method could allow more flexible application of VM migration without requiring any change in the terminal environment. The image of the proposed method is illustrated in Figure 1.

## 2.2 Method of dynamically inserting WAN accelerator after migration of virtual machines

Even when the IP addresses of VMs are changed as a result of wide-area live migration, it could be possible to maintain communication using the methods proposed in References [24]-[28]. Those methods do not support the establishment of a new TCP connection in conjunction with VM migration, even though most of existing WAN accelerators initiate WAN optimization function by establishing a new TCP connection. Therefore, we have chosen to use mSCTP[29]-[31], which supports multi-homing and multiple IP addresses simultaneously.

In mSCTP-based migration, VMs will transfer data using different TCP connections before and after migration, thereby maintaining the sessions between the terminals and servers. Issues form SCTP-based live migration and solutions to them are described and proposed below. It is assumed here that VirtualBox [32] is used as the server virtualization system. The 'teleportion' function of VirtualBox supports live migration.

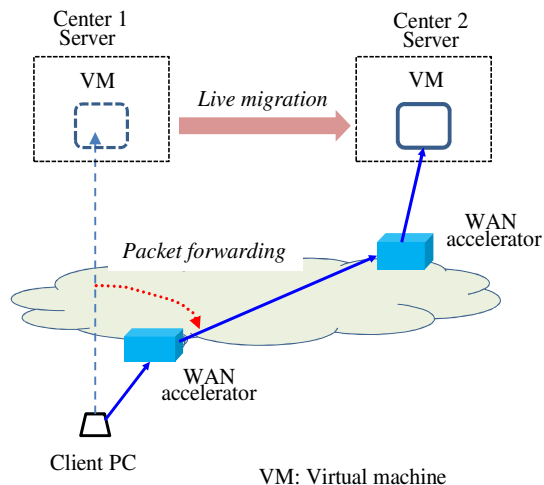

Figure 1.  Image of wide-area live migration of VM with WAN optimization function

(1) How client PC finds out the new IP address of the destination VM

It is impossible to know in advance when VM migration will occur. It is proposed that the source server notifies the client terminal of the new IP address of the destination VM using SCTP messages in advance.

(2) When and how to determine migration of VMs

It is generally difficult for an external party to know immediately the timing when VM

43

migration has completed. It is proposed to solve this problem as follows: After VM migration has been completed, VirtualBox's window of the destination server changes from the migration standby & display window to a VM's OS activation window. Completion of VM migration can be recognized by watching for this change. The client PC is notified of this change separately from the mSCTP functions. The reception of this notification prompts the client PC to request the migrated VM on the destination server to establish a new TCP connection.

Figure 2 shows the communication sequence of the entire mSCTP-based live migration that incorporates the solutions described in (1) and (2). The image of connection management in client PCs and servers is illustrated in Figure 3. Each client PC and VM has two different IP addresses. It uses different IP addresses and uses different TCP connections before and after VM migration. The transferred data is first stored in SCTP packets, and is then encapsulated in TCP packets and transmitted.

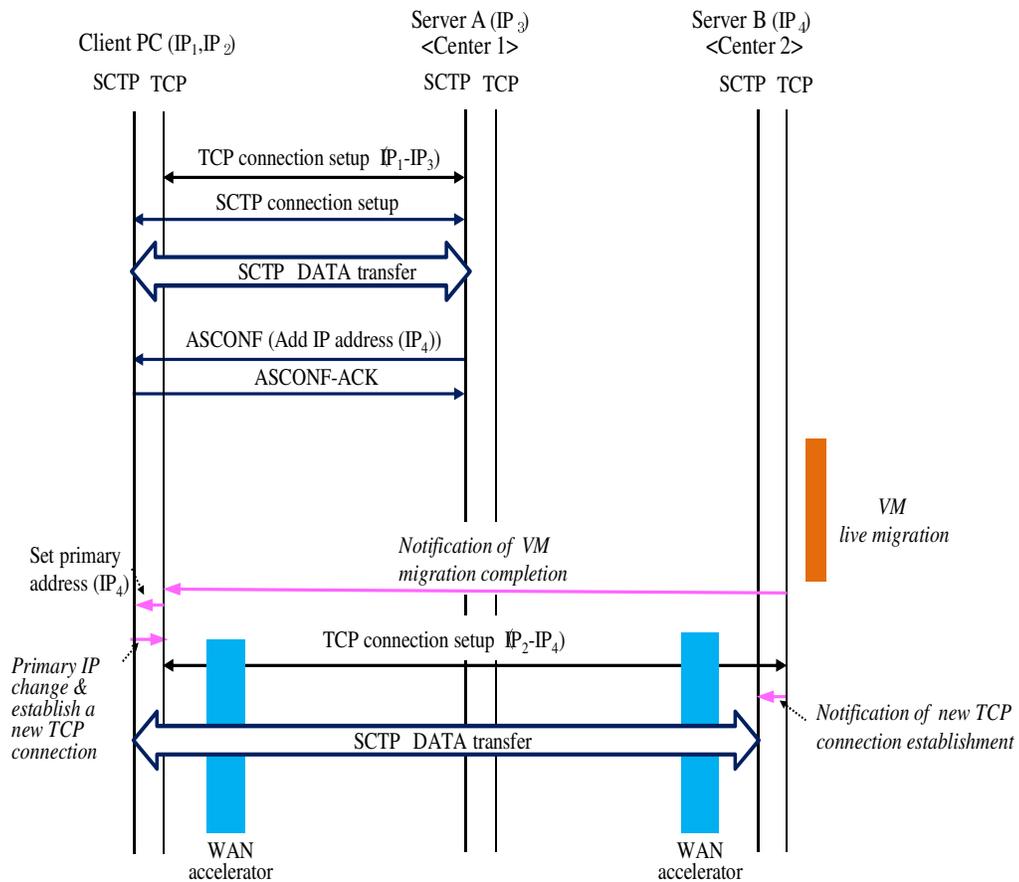

Figure 2. Overview of VM migration sequence with mSCTP

## 2.3 Evaluation of mSCTP-based live migration

### 2.3.1 Evaluation system configuration

A system shown in Figure 4 was built to verify the operation of the proposed method. The specification of individual devices in the system is as follows:

<Servers>
  CPU: Intel Core™ i5, Virtualization system: VirtualBox 4.0.16, Host OS:  Windows 7, Guest OS: Windows XP, LAN interface: 100Mbps
<Clients>
  CPU: Intel Core™ i5, OS:  Windows 7, LAN interface: 100Mbps
＜NAS＞
  1 TB×2, LAN interface: 100Mbps
＜Network Emulator＞
  Maximum speed: 100Mbps, constant delay
<WAN accelerator>

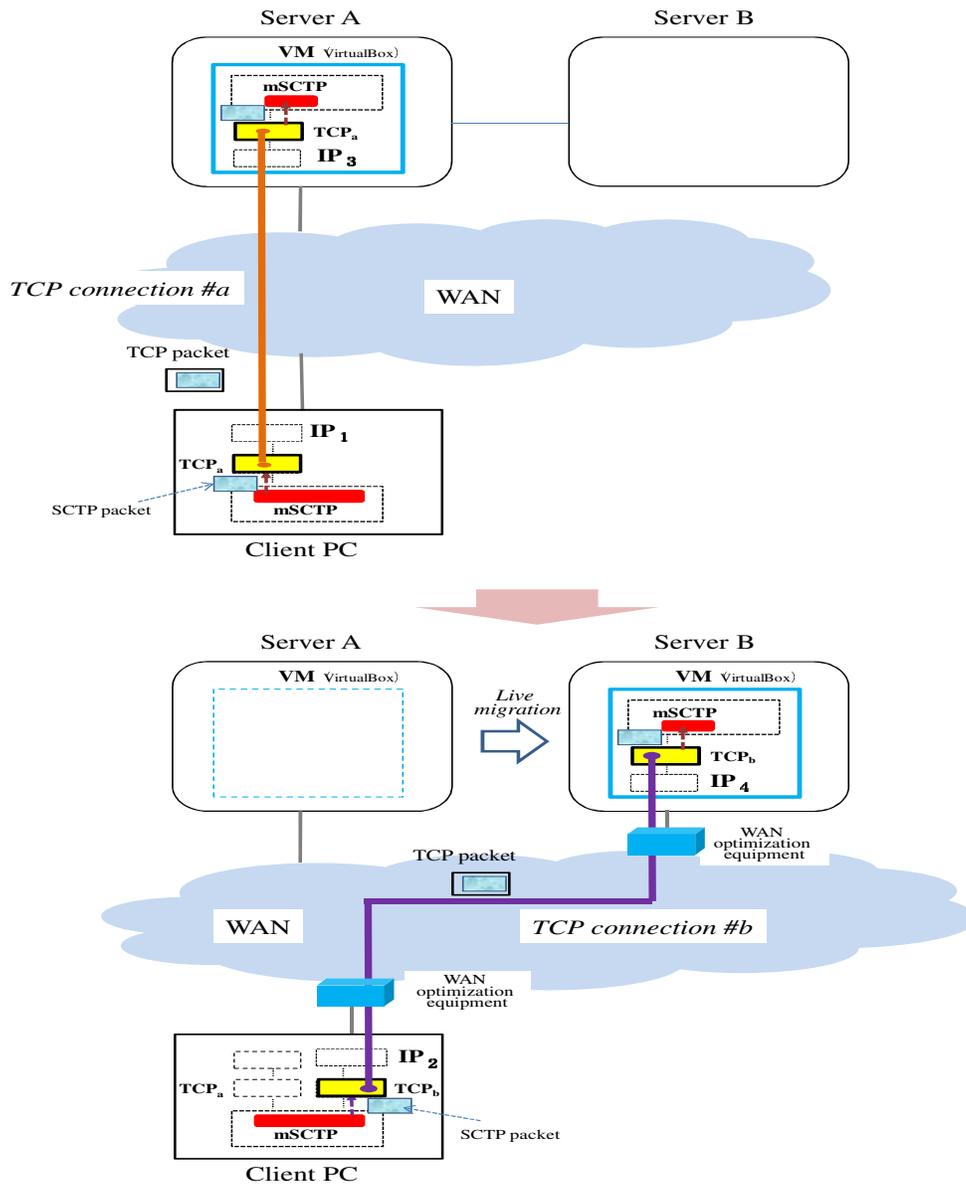

Figure 3.  Image of TCP connection exchange in servers and client PC

Steelhead （Riverbed） Maximum throughput: 1Mbps

### 2.3.2 Verification of the operation

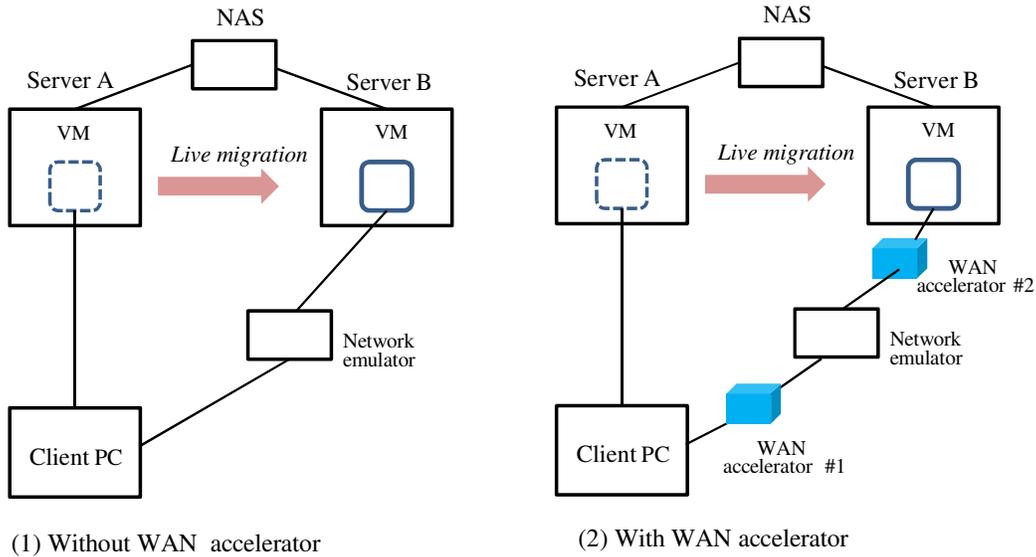

(1) Without WAN accelerator

(2) With WAN accelerator

Figure 4. Evaluation system configuration

We have checked the live migration (teleportation) of a VM on Server A to Server B in Figure 4 while the VM is gradually displaying an image, up to the point where the entire image has been displayed. It was assumed that no other communication is in progress. We have found that the image displaying speed is reduced to one-fifth of the speed before the VM migration unless a WAN accelerator is inserted, when the network delay between Server B and the client PC is 300 ms. The insertion of a WAN accelerator after the migration returns the image displaying speed to almost the same speed as in the pre-migration state. It has been also found that the insertion of a WAN accelerator has little impact on VM migration time.

Table 1 shows the time it takes to transfer file data with FTP. d is the network delay between Server B and the client PC and is assumed to be constant. T shows the time which is taken to complete the transfer of all data(50MB) after live migration with WAN accelerator and does not include the time required for live-migration. The value of T is normalized relative to the time required to complete the transfer of all data without live migration. It is clear from Table 1 that WAN accelerator makes it possible to maintain the communication performance similar to that before the migration.

Table 1. Required time to complete the transfer of all data with WAN accelerator

| d [ms] | T |
|---|---|
| 250 | 0.25 |
| 500 | 0.1 |
| 750 | 0.08 |
| 1000 | 0.07 |

*d: Network delay between client and server B after live migration*
*T: Required time to complete the transfer of all data (50MB) after live migration which is normalized relative to the time without WAN accelerator*

## 3. Method of reducing power consumption of ICT devices

### 3.1 Reducing power consumption by applying WAN accelerator[18]

In recent years, enterprises are consolidating servers and storage units in different sites into a single center, in order to utilize those resources efficiently and reduce their business cost. However, the benefit of efficient operation may derive at the expense of reduced performance and inconvenience for users since the distance to those resources becomes longer than before. To prevent these negative effects, WAN accelerators[14]-[17] are increasingly introduced. In TCP, ACK proxy responses at the WAN accelerator side, widening of the TCP window size, etc. can shorten response time and increase throughput. In addition, data compression and caching can reduce traffic in the WAN, thereby preventing congestion and reducing the required bandwidth.

In addition, WAN accelerators can reduce the power consumption of ICT devices. That is, WAN accelerators can dramatically shorten communication time in a file transfer application and some ICT devices can be put in sleep mode for a long time by the shortened time. It can reduce the power consumption of them. This idea could not apply to all applications but is effective for huge data transfer.

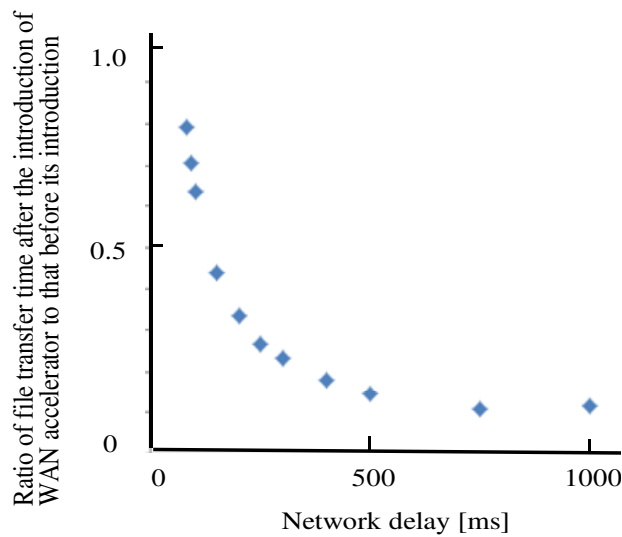

Figure 5. Effect of introducing WAN accelerator

Figure 5 illustrates an example of measurement of time it takes to transfer a fixed-size data file with FTP. The same WAN accelerator (Steelhead) in Section 2.3.1 is also used here. Network delay is assumed to be constant. It shows how the introduction of a WAN accelerator shortens communication time. The vertical axis indicates the ratio of the file transfer time after the introduction of a WAN accelerator to that before its introduction. For example, where the network delay is 500ms, the introduction of WAN accelerator can shorten file transfer time to about 1/10 and thereby the power consumed by the ICT devices connected to the WAN would be reduced by 1/10 at the maximum.

While the individual user can introduce a WAN accelerator on their own, it is effective to install WAN accelerators as part of cloud resources and lease them to users on an hourly basis, as proposed in Reference [17], or for a network provider to apply WAN accelerator transparently. If WAN accelerators are combined with other WAN optimization technologies, such as data compression and caching, it is possible to reduce not only power consumption but also network bandwidth that the network provider needs to install.

## 3.2 Reducing power consumption by increasing the packet transfer rate

As a measure to reduce the power consumption of the network, it has been proposed to reduce the packet transfer rates of communication links when traffic on these links is small [33]-[35].  This idea assumes that the higher the link transfer rate, the more power is consumed. However, in applications, such as file transfer, that transfer a huge volume of data continuously but do not require real-time transfer, the total power consumption can be reduced by increasing the packet transfer rate conversely, thereby shortening the communication time.

Let $T_h$ be the communication time in the case where fast transfer is used, $P_h$ be the electric power of the communication link in the same case, $T_l$ be the communication time in the case where slow transfer is used, and $P_l$ be the electric power of the communication link in the same case.  As the inequality, $T_h/T_l > P_h/P_l$, generally holds, the comparison of the two cases in the total power consumption becomes $T_h*P_h < T_l*P_l$. That is, the total power consumption will be smaller when data transfer is faster. In addition, the faster transfer makes it possible not only to reduce the total power consumption of the communications link but also to put the link in sleep mode for a longer time than when slower transfer is used. This can further reduce the power consumption of the link. Furthermore, the higher the probability at which all links connected to a node are put to sleep mode, the higher the probability at which the node itself can be put to sleep mode.

Figure 6 illustrates one example of total power consumption of communication switch which compares the case where a file data is transferred at a constant low speed with the case where a file data is transferred at a high speed. The fixed part in the figure is the total power consumption except NICs. The values of electric power are assumed to be the same as those of our evaluation system. It is also supposed here that the devices are put to sleep after the completion of the transfer. In this example, the power consumption with high speed is only about one-tenth of that with low speed.

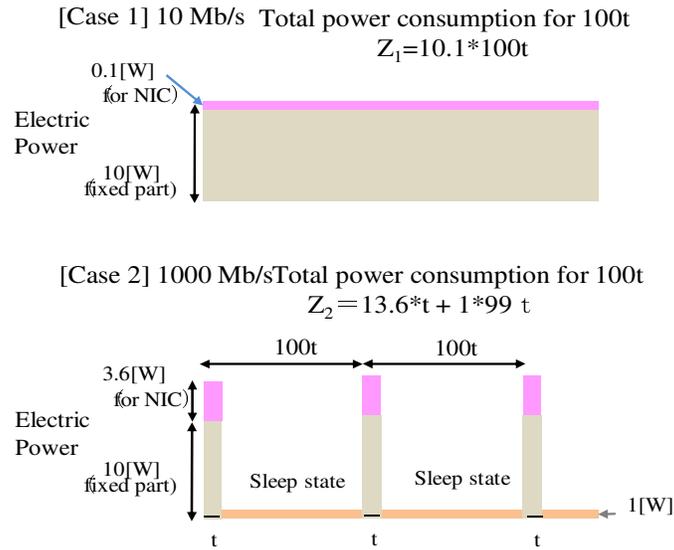

Figure 6. Comparison of total power consumption by link speed

## 4. Related work

A variety of solutions such as Mobile IP [24], SIP mobility [25], TCP migration [26], LISPmob [27], mSCTP [29]-[31]could be applied to mobility management on VM live migration over a wide area.  As it is required to establish a new TCP connection to initiate WAN optimization function supported by the existing WAN accelerators, we proposed to adopt mSCTP which supports multiple IP addresses simultaneously in Section 2. Moreover, multiple methods have been proposed to improve or to prevent the degradation of the performance of wide-area live migration itself [4]-[13]. However, most of them do not consider to prevent the degradation in performance after live migration of VM over a wide area.

WAN accelerators have been introduced to prevent degradation in performance caused by a long network delay in WANs originally. They can also dramatically shorten communication time of applications such a file transfer and thereby reduce the power used by data centers and network devices, as proposed in Section 3.1.

As for a measure to reduce the power consumed by the network, Adaptive Link Rate (ALR) technology has been proposed [33]-[35]. This idea assumes that the higher the link transfer rate, the more power is consumed. However, in applications, such as file transfer, that transfer a huge volume of data continuously but do not require real-time transfer, the total power consumption can be reduced by increasing the packet transfer rate conversely, thereby reducing the packet transfer rates of communication links when traffic on these links is small, as proposed in Section 3.2

International Journal of Computer Networks & Communications (IJCNC) Vol.5, No.1, January 2013

## 5. Conclusions

This paper has proposed to dynamically apply WAN accelerator within the network when a virtual machine is moved to a distant center, in order to prevent the degradation in performance after live migration of virtual machines over a wide area. mSCTP-based data transfer using different TCP connections before and after migration have been proposed in order to use a currently available WAN accelerator.  Assuming VirtualBox for a virtualization system, we have verified the operation of the proposed method, and confirmed that it is possible to prevent degradation in communication performance after migration of virtual machines. This paper has not considered the performance degradation of live migration itself.

Then, this paper has proposed the method to reduce the power consumption of ICT devices in a cloud computing environment, which consists of installing WAN accelerators as part of cloud resources actively and increasing the packet transfer rate of communication link temporarily.   It has been indicated that the power consumption with WAN accelerator could be reduced to one-tenth of that without WAN accelerator.

It is necessary to study more specific schemes for the installation of WAN accelerators in a network, conditions that the introduction of WAN accelerator is effective, and details of the operation of WAN accelerators on VM live migration. It is also necessary to study the detailed schemes and required protocols for the installation of WAN accelerator to reduce the power consumption of ICT devices in cloud computing environments.

## Acknowledgement

xThis work was partly supported by MEXT (Japan) grant-in-aid for building strategic research infrastructures.

## References

x

## Author

**Shin-ichi Kuribayashi** received the B.E., M.E., and D.E. degrees from Tohoku University, Japan, in 1978, 1980, and 1988 respectively. He joined NTT Electrical Communications Labs in 1980.  He has been engaged in the design and development of DDX and ISDN packet switching, ATM, PHS, and IMT 2000 and IP-VPN systems.  He researched distributed communication systems at Stanford University from December 1988 through December 1989. He participated in international standardization on ATM signaling and IMT2000 signaling protocols at ITU-T SG11 from 1990 through 2000.  Since April 2004, he has been a Professor in the Department of Computer and Information Science, Faculty of Science and Technology, Seikei University.  His research interests include optimal resource management, QoS control, traffic control for cloud computing environments and green network.  He is a member of IEEE, IEICE and IPSJ.

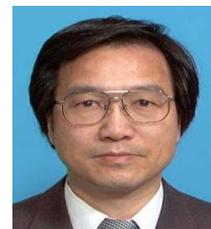